\documentstyle[editedvolume, numreferences]{crckapb} 

\begin{opening}
\title{CHIRAL SYMMETRY BREAKING IN STRONGLY COUPLED \protect $1+1$ DIMENSIONAL LATTICE GAUGE THEORIES${}^1$}
\author{F. Berruto, G. Grignani and P. Sodano}
\institute{Dipartimento di Fisica and Sezione I.N.F.N. , \\Universit\'a di 
Perugia,\\Via A. Pascoli I-06123, Perugia, Italy}

\end{opening}

\runningtitle{CHYRAL SYMMETRY BREAKING}

\begin{document}

\section{Abstract}
We analyse $-$ within the hamiltonian formalism with staggered fermions $-$ 
the patterns of chiral symmetry breaking for the strongly coupled Schwinger 
and $U({\cal N}_c)$-color `t Hooft models with one and two flavor of fermions. 
Using the correspondence between these strongly coupled gauge models  and 
antiferromagnetic spin chains, we provide a rather intuitive picture of their 
ground states, elucidate their patterns of chiral symmetry breaking, and compute the pertinent 
chiral condensates. Our analysis evidences an intriguing 
relationship between the values of the lattice chiral condensates of the 
`t Hooft and Schwinger models with one flavor of fermions.

\section{Introduction}
\addtocounter{footnote}{1}
\footnotetext{Contribution to the proceedings of the workshop 
``Lattice fermions and the structure of the vacuum'', 5-9 October 1999, Dubna, Russia.}
Confinement and chiral symmetry breaking go hand in hand as strong coupling 
phenomena in a gauge theory; while confinement is an observed property of the
strong interactions and it is an unproven, but widely believed feature of 
most non-Abelian gauge theories in four and lower space-time dimensions, 
chiral symmetry is only an approximate symmetry of particle physics 
since the up and down quarks are light but not massless. 

The strong coupling limit of lattice gauge theories, even if far from the 
scaling regime, is useful to study interesting properties of a gauge 
model. In this limit, the gauge theory exhibits confinement and dynamical 
chiral symmetry breaking: Wilson fermions explicitly break chiral symmetry 
while for staggered fermions, even if the continuous chiral symmetry is 
broken explicitly, a discrete axial symmetry survives the lattice 
regularization and it is realized on the lattice theory as a translation by 
one site. 

There is an interesting issue arising in 
the lattice regularization of gauge theories, which  is how the lattice 
theory produces the effects of the axial anomaly. 
Usually, on the lattice axial anomalies are either cancelled by fermion 
doubling or else the lattice regularization breaks the axial symmetry 
explicitly. Lattice gauge theories in $1+1$ dimensions, in the Hamiltonian 
formalism with staggered 
fermions, represent the unique example where neither of this occurs.
For lattice models in higher space-time dimensions or for a full lattice 
definition of non-anomalous chiral gauge theories one should resort to the 
more powerful overlap construction~\cite{neuberger} of lattice fermions, 
which preserve 
global chiral symmetries on the lattice in theories such as 
$QCD$~\cite{narayanan}.
 
For the $(1+1)$-dimensional lattice theories the effects of the anomaly are 
not canceled by doubling since 
the continuum limit of a $(1+1)$-dimensional staggered fermion produces 
exactly one Dirac fermion; moreover, even if the 
continuum axial symmetry is explicitly broken by staggered fermions, 
a discrete axial symmetry survives on the lattice. 
It corresponds to the continuum transformation
\begin{equation}
\psi(x)\longrightarrow \gamma^5 \psi(x)\quad,\quad \overline{\psi}(x)\longrightarrow - \overline{\psi}(x) \gamma^5
\label{das}
\end{equation}
with $\gamma^5=\sigma^1$; it is realized on the lattice as a translation by 
one site. 
Since the mass operator $\overline{\psi}\psi$ is odd under 
Eq.(\ref{das}), if this symmetry is unbroken 
then $\langle \overline{\psi}\psi \rangle=0$. This happens for both the 
`t Hooft and Schwinger models with two flavors of fermions
~\cite{berruto1,berruto2}. 

For the one-flavor models the discrete axial symmetry should be broken
~\cite{berruto2,berruto3} since the one-flavor continuum 
Schwinger model has a non-zero chiral condensate given by~\cite{swieca}
\begin{equation}
\langle \overline{\psi}(x)\psi(x) \rangle =-\frac{e^{\gamma}}{2\pi}\frac{e_c}{\sqrt{\pi}}\quad ,
\label{1scc}
\end{equation}
with $\gamma=0.577\ldots$ the Euler constant and the one-flavor `t Hooft model exhibits a non-zero chiral condensate 
given by~\cite{zhitnitsky}
\begin{equation}
\langle \overline{\psi}\psi \rangle=-{\cal N}_{c}(\frac{g_c^{2}{\cal N}_{c}}{12\pi})^{\frac{1}{2}}\quad .
\label{1tcc}
\end{equation}  
In the following we shall discuss the pattern of chiral symmetry breaking 
and review the lattice computation of the pertinent chiral condensates 
for the Schwinger and `t Hooft models with one and two flavors of 
fermions~\cite{berruto1,berruto2,berruto3}.

Staggered fermions are a useful tool to investigate $1+1$ dimensional 
lattice gauge theories also because  
$-$ in the strong coupling limit $-$ they provide an explicit correspondence 
between the gauge theory and a pertinent spin system~\cite{semenoff}. 
The mapping is useful since it provides not only  an intuitive picture of 
the ground state of the 
gauge model but also the mass spectrum and the lattice chiral 
condensate of the gauge model in terms of 
spin correlators of the 
corresponding spin models~\cite{berruto1,berruto2}. 
The relevant differences between the one- and multi-flavor gauge models may 
be intuitively represented in terms of this 
correspondence: the two-flavor Schwinger and `t Hooft models 
correspond to the physically 
relevant $SU(2)$ quantum Heisenberg 
antiferromagnets~\cite{berruto1,berruto2}, while the one-flavor models 
correspond to the Ising antiferromagnets. 
The ground states of these spin models are very different: while the Ising 
antiferromagnet exhibits 
spontaneous breaking of the discrete axial symmetry, this 
does not happen for the quantum $SU(2)$ 
Heisenberg antiferromagnet. 
The order parameters signaling chiral symmetry breaking are 
either an isoscalar chiral condensate, $\chi_{isos.}=\langle \overline{\psi}\psi \rangle$, or an isovector chiral condensate 
$\chi_{isov.}=\langle \overline{\psi}\sigma^a \psi\rangle$: they can be non-vanishing only for the one-flavor models. 
For the two-flavor models the only relic~\cite{berruto1,berruto2,gattringer} 
of chiral symmetry breaking is the non-vanishing of $\langle \overline{\psi^2}_L \overline{\psi^1}_L\psi^1_R\psi^2_R\rangle$.

\section{One-flavor Schwinger and `t Hooft model chiral condensates}

In this section we briefly review the lattice 
strong coupling computation of the chiral condensates for the one-flavor 
Schwinger and `t Hooft 
model~\cite{berruto2,berruto3}.
 
For both models the strong coupling Hamiltonian may be presented as $H=H_0+\epsilon H_h$, where $H_0=\sum_{x}(E_x)^2$ is the 
unperturbed hamiltonian and $H_h=-i(R-L)$ is the hopping 
Hamiltonian, with 
$R=\sum_x\psi_{a x+1}^{\dagger}U_{ab}(x)\psi_{bx}$ the right hopping operator ($L=R^{\dagger}$). $U_{ab}(x)$ is a matrix 
defined on the link $[x,x+1]$ in the non-Abelian model~\cite{berruto2} 
or a phase in the Abelian model~\cite{berruto1,berruto3} and $\epsilon=t/g^2a^2$ 
is the strong coupling expansion parameter, 
with $g$ the coupling constant and $a$ the lattice spacing. 

In the strong coupling limit one has to find states which are annihilated by 
the generator of gauge transformations and at the 
same time are eigenstates of the unperturbed Hamiltonian $H_0$. $H_0$ exhibits two degenerate ground states, 
$|g.s.\rangle_1$ characterized by a 
charge distribution on each even site of one (${\cal N}_c$) particle(s) in the Abelian(non-Abelian) model, and 
$|g.s.\rangle_2$ characterized by the same charge distribution on each odd site. 
Each of these ground states spontaneously breaks the discrete axial 
symmetry (\ref{das}), since by translating by one lattice 
spacing one ground state one gets to the other one. 
The thermodynamic limit is taken so that these two states 
are not mixed to any finite order of perturbation theory~\cite{berruto3}. 
Consequently, one should carry out non-degenerate perturbation theory only 
around one ground state which we shall denote by $|g.s.\rangle$: one has spontaneous breaking of the discrete axial symmetry. 
At the second order in the strong coupling expansion both models are 
effectively described by the antiferromagnetic Ising model, 
whose ground state is the classical N\'eel configuration $|g.s.\rangle$~\cite{berruto2}.

We shall now show that the effects of the dynamical symmetry breaking, 
due to the anomaly in the continuum models, is reproduced on the lattice 
through the breaking of the discrete axial symmetry. 
For this purpose one should verify that the chiral condensate is 
non-vanishing. 
In the staggered fermion formalism, the lattice chiral condensate may be 
obtained by computing the v.e.v. of the mass operator 
$M=-1/Na \sum_{x=1}^{N}\sum_{a=1}^{{\cal N}_c}(-1)^x \psi^{\dagger}_{ax}\psi_{ax}$ on the perturbed states $|p_{g.s.}\rangle$ generated by applying 
$H_{h}$ to $|g.s.\rangle$. 
To the second order in $\epsilon$, $|p_{g.s.}\rangle$ 
is given by
\begin{equation}
|p_{g.s.}\rangle=|g.s.\rangle+\epsilon|p_{g.s.}^{(1)}\rangle+\epsilon^2 |p_{g.s.}^{(2)}\rangle \quad ,
\end{equation} 
with
\begin{eqnarray}
|p_{g.s.}^{(1)}\rangle&=&\frac{\Pi_{g.s.}}{E_{g.s.}^{(0)}-H_{0}}H_{h}|g.s.\rangle\quad ,\\
|p_{g.s.}^{(2)}\rangle&=&\frac{\Pi_{g.s.}}{E_{g.s.}^{(0)}-H_{0}}H_{h}\frac{\Pi_{g.s.}}{E_{g.s.}^{(0)}-H_{0}}H_{h}|g.s.\rangle\quad ,
\end{eqnarray}
and $E_{g.s.}^{(0)}$ is the $|g.s.\rangle$ eigenvalue of $H_0$. 
$\Pi_{g.s.}$ is a projector orthogonal to $|g.s.\rangle$.

To the fourth order in $\epsilon$, the lattice chiral condensate 
is then given by 
\begin{eqnarray}
\chi_{L}&=&\frac{\langle p_{g.s.}|M|p_{g.s.}\rangle}{\langle p_{g.s.}|p_{g.s.}\rangle}\nonumber\\
&=&\frac{\langle g.s.|M|g.s.\rangle+\epsilon^2 \langle p_{g.s.}^{(1)}|M|p_{g.s.}^{(1)}\rangle+\epsilon^4 \langle p_{g.s.}^{(2)}|M|p_{g.s.}^{(2)}\rangle}
{\langle g.s.|g.s.\rangle+\epsilon^2 \langle p_{g.s.}^{(1)}|p_{g.s.}^{(1)}\rangle+\epsilon^4 \langle p_{g.s.}^{(2)}|p_{g.s.}^{(2)}\rangle}\quad .
\label{1c}
\end{eqnarray}

By direct computation \cite{berruto3} in the Schwinger model 
one gets
\begin{equation}
\chi^{Schwinger}_{L}=-\frac{1}{a}(\frac{1}{2}-8\epsilon^2 +96\epsilon^4)\quad .
\label{l1scc}
\end{equation}
In the `t Hooft model one finds a nonvanishing chiral condensate for any 
finite 
${\cal N}_c$~\cite{berruto2}, in agreement with the results of 
Ref.~\cite{grandou}, and in the limit 
${\cal N}_c\longrightarrow \infty$ one gets 
\begin{equation}
\chi^{`t\ Hooft}_{L}=-\frac{1}{a}{\cal N}_{c}(\frac{1}{2}-8\epsilon^2 +32\epsilon^4)\quad .
\label{l1tcc}
\end{equation}
The results given in Eqs.(\ref{l1scc},\ref{l1tcc}) well agree 
with the continuum answers~\cite{berruto1,berruto2}.

We observe~\cite{berruto2} that the lattice computation of the chiral condensate shows that, 
up to order $\epsilon^2$, the chiral condensate given in Eq.(\ref{l1tcc}) is 
just ${\cal N}_c$ times 
the chiral condensate given in Eq.(\ref{l1scc}) for the one-flavor Schwinger 
model. 
Taking into account the contributions up to the fourth order in $\epsilon$ 
one has
\begin{equation}
\chi_{L}^{`t\ Hooft}={\cal N}_{c}(\chi_{L}^{Schwinger}+\frac{1}{a}64\epsilon^4)\quad .
\label{chichi}
\end{equation}
The difference is due to terms such as $\langle g.s.|\sum_{x=1}^NR_xR_xL_xL_x|g.s.\rangle$, 
which are non vanishing only in the non-Abelian model, since in the Abelian 
model the sites of 
$|g.s.\rangle$ are either empty or occupied by just one particle: 
$|g.s.\rangle$ is annihilated 
when more than one hopping operator acts on the same site. 
The lattice computation, which brought us to Eq.(\ref{chichi}), shows that 
the chiral condensate of the `t Hooft model factorizes to all orders in the product of 
${\cal N}_c$ and one term whose numerical value is dominated by the 
contribution of the chiral condensate of the one-flavor Schwinger model; 
as seen from Eq.(\ref{chichi}) the leading correction to the Abelian chiral condensate is of order $O(\epsilon^4)$. 
>From Eq.(\ref{chichi}) one may be lead to conjecture that the chiral condensate of the non-Abelian model is determined mainly by 
the $U(1)$ Abelian subgroup of $U({\cal N}_{c})$; the non-Abelian group 
contributing a factor proportional to ${\cal N}_c$.  

\section{Chiral symmetry breaking in the two-flavor Schwinger and 
`t Hooft models}

The two-flavor Schwinger and `t Hooft models are effectively described 
$-$ at the second order in the strong coupling expansion $-$  
by a quantum antiferromagnetic Heisenberg chain of spin-$1/2$ and spin-${\cal N}_c/2$, respectively~\cite{berruto1,berruto2}. 
The vacuum of the gauge models, $|g.s.\rangle$, becomes the ground state of 
the spin models in the infinite coupling limit; with periodic boundary 
conditions this state is translationally invariant and, consequently,  
the discrete lattice chiral symmetry is not 
broken when there are two flavors of fermions. 

The translational invariance of $|g.s.\rangle$ implies that both 
$\chi_{isos.}$ and $\chi_{isov.}$ are zero to all the orders in the strong 
coupling expansion. If one introduces~\cite{berruto1} 
the translation operator 
$\hat{T}=e^{i \hat{p} a}$ and takes into account that 
\begin{eqnarray}
\hat{T}M\hat{T}^{-1}&=&-M\quad ,\\
\hat{T}\Sigma \hat{T}^{-1}&=&-\Sigma\quad ,\\
\hat{T}H_{h}\hat{T}^{-1}&=&H_{h}\quad ,\\
\hat{T}|g.s.>&=&\pm|g.s.>
\end{eqnarray}
with 
\begin{equation}
M=(1/2Na)\sum_{a=1}^{{\cal N}_c}\sum_{\alpha=1}^2
\sum_{x=1}^N(-1)^x \psi_{ax}^{\alpha\ \dagger}\psi_{ax}^{\alpha}
\end{equation} 
and 
\begin{equation}
\vec{\Sigma}=(1/2Na)\sum_{x=1}^{N}(-1)^x\psi_{a,x}^{\dagger}
\vec{\sigma}_{ab}\psi_{b,x}\quad ,
\end{equation}
one gets that $\chi_{isos.}=-\chi_{isos.}$ 
and $\chi_{isov.}=-\chi_{isov.}$ to all the orders in the strong coupling expansion. In the continuum these order parameters are naturally zero since they 
would signal the breaking not only of the $U_A(1)$ symmetry of the action but 
also of the internal flavor symmetry $SU_L(2)\otimes SU_R(2)$, which 
is protected by Coleman's theorem~\cite{coleman}. 

The nonvanishing v.e.v. signaling the breaking of only the 
$U_A(1)$ chiral symmetry is 
$\langle F \rangle=\langle \overline{\psi^{2}}_L \overline{\psi^{1}}_L \psi_{R}^{1} \psi_R^{2} \rangle$.
On the lattice, $F$ is written as~\cite{berruto1}
\begin{equation}
F=-\frac{1}{16a^2 N} \sum_{x=1}^N \left\{ (n_x^1-n_{x+1}^1)(n_x^2-n_{x+1}^2)+(L_x^1-R_x^1)(L_x^2-R_x^2) \right\}\quad .
\label{ff}
\end{equation}
In Eq.(\ref{ff}) $n_x^{\alpha}=\sum_{a=1}^{{\cal N}_c}
\psi_{ax}^{\alpha \dagger}\psi_{ax}^{\alpha}$ are the occupation numbers at site $x$.
The strong coupling expansion, up to the second order in $\epsilon$, 
yields 
\begin{equation}
\langle F \rangle=\frac{\langle p_{g.s.}|F|p_{g.s.}\rangle}{\langle p_{g.s.}|p_{g.s.}\rangle}=
\frac{\langle g.s.|F|g.s.\rangle +\epsilon^2 \langle p_{g.s.}^{(1)}|F|p_{g.s.}^{(1)}\rangle }
{\langle g.s.|g.s.\rangle +\epsilon^2 \langle p_{g.s.}^{(1)}|p_{g.s.}^{(1)}\rangle }\quad .
\label{cc2}
\end{equation}

For both the two-flavor Schwinger and `t Hooft models, it is possible to 
explicitly compute $\langle F\rangle$ 
in terms of spin correlators of the 
antiferromagnetic spin-$1/2$ and spin-${\cal N}_c/2$ models, respectively.
For the Abelian model the wave function are normalized as
\begin{eqnarray}
<g.s.|g.s.>&=&1\\
<p_{g.s.}^{1}|p_{g.s.}^{1}>&=&-4<g.s.|H_{J}|g.s.>\label{1e}\quad ,
\end{eqnarray}
and taking into account that
\begin{eqnarray}
<g.s.|F|g.s.>&=&\frac{1}{8a^{2} N}<g.s.|H_{J}|g.s.>\label{2e}\\
<p^{1}_{g.s.}|F|p^{1}_{g.s.}>&=&
\frac{1}{4a^{2}N}(-2<g.s.|( H_{J})^{2}|g.s.>-\frac{5}{3}
<g.s.|H_{J}|g.s.>\nonumber \\
&+&\frac{5}{12}N-\frac{2}{3}\sum_{x=1}^{N}<g.s.|\vec{S}_{x}\cdot \vec{S}_{x+2}-
\frac{1}{4}|g.s.> )\label{3e}\quad ,
\end{eqnarray}
from Eqs.(\ref{cc2}), using the known correlation functions of the Heisenberg model to 
numerically evaluate the v.e.v.'s given in Eqs.(\ref{1e},\ref{2e},\ref{3e}), one gets 
\begin{equation}
<F>=\frac{1}{a^{2}}(0.0866-0.4043\epsilon ^{2})\quad .
\label{nopuad}
\end{equation}
In the non-Abelian model the wave functions are normalized as 
\begin{eqnarray}
\langle g.s.|g.s. \rangle &=& 1\quad , \\
\langle p_{g.s.}^{(1)} |p_{g.s.}^{(1)} \rangle &=&-\frac{16}{{\cal N}_c}\langle g.s.|H_J|g.s.\rangle\quad ,\label{ee1} 
\end{eqnarray}
and the equations corresponding to Eqs.(\ref{2e},\ref{3e}) are now given by
\begin{eqnarray}
\langle g.s.|F|g.s.\rangle &=& -\frac{1}{16a^2N}\left(\frac{2}{3}\langle g.s.|H_J|g.s. \rangle -\frac{{\cal N}_c^2}{3}N\right)
\quad ,\label{ee2}\\
\langle p_{g.s.}^{(1)}|F|p_{g.s.}^{(1)} \rangle &=& -\frac{1}{16a^2N} 
\left[ \frac{32}{3{\cal N}_c}\langle g.s.|H_J^2|g.s. \rangle
+\frac{16}{3}({\cal N}_cN-1)\langle g.s.|H_J|g.s. \rangle \right. \nonumber \\ 
& &+8{\cal N}_c^2N+\frac{16}{3}\langle g.s.|\sum_{x=1}^{N}(\vec{S}_x\cdot \vec{S}_{x+2}-
\frac{{\cal N}_c^2}{4})|g.s.\rangle \nonumber \\
& &+\left. \frac{64}{3{\cal N}_c^2}\sum_{x=1}^{N}\langle g.s.|(\vec{S}_{x}\cdot \vec{S}_{x+1})^2|g.s.\rangle \right] \quad .\label{ee3}
\end{eqnarray}

Since for spins higher than $1/2$, no analytical result concerning 
correlation functions is known, one has to evaluate 
Eqs.(\ref{ee1},\ref{ee2},\ref{ee3}) in the large 
spin limit $S\rightarrow \infty$, which $-$ since 
$S={\cal N}_c/2$ $-$ corresponds to the planar limit 
${\cal N}_c\rightarrow \infty$ of the gauge theory~\cite{thooft}.
>From Eq.(\ref{cc2}) one then gets
\begin{equation} 
\langle F \rangle=\frac{{\cal N}_c^2}{a^2}\left(0.042-0.750\epsilon^2 \right)
\quad .
\label{ccc2}
\end{equation}
The nonvanishing v.e.v. determined by Eq.(\ref{ccc2}) is the lattice relic of the $U_A(1)$ anomaly in the continuum theory. 
As evidenced in Refs.~\cite{berruto1,berruto2} the operator $F$ 
describes, on the lattice, an umklapp process.

\section{Concluding remarks}

We reviewed the analysis of the chiral symmetry breaking patterns in 
strongly coupled $1+1$ dimensional gauge theories such as the Schwinger and `t Hooft models on the 
lattice~\cite{berruto1,berruto2,berruto3}; since the 
``doubling''  of fermion species is 
completely removed by staggered fermions in $1+1$ dimensions one expects that
the lattice 
regularization faithfully reproduces in these cases the results of the 
continuum theory. 

Using the correspondence 
between the strongly coupled lattice Schwinger and `t Hooft models with 
antiferromagnetic spin chains derived in 
Refs.~\cite{berruto1,berruto2,berruto3}, one has that, while 
the one-flavor models are effectively described by antiferromagnetic Ising 
chains, the two-flavor 
models are effectively described by antiferromagnetic Heisenberg chains. 
This correspondence is useful in providing not only a rather 
intuitive picture of the ground state of a gauge model $-$ and thus of the 
patterns of chiral symmetry breaking on the lattice$-$ but also $-$ as evidenced in Section 4 $-$ an 
expression for the chiral condensates in terms of spin correlators of 
the pertinent Heisenberg chain. 

It would be interesting to exhibit the mapping of gauge theories onto 
antiferromagnetic Heisenberg models also 
in the context of the ``overlap'' fermions~\cite{neuberger}. 
This correspondence is known to exist in all the 
previously known approaches to lattice fermion, as evidenced in 
Ref.~\cite{weinstein1} using the SLAC derivative, in 
Ref.~\cite{smit} in the context of Wilson fermions and in 
Refs.~\cite{semenoff,berruto1} using staggered fermions. Moreover, 
an intriguing question to answer is if the mapping of gauge theories onto 
quantum antiferromagnets survives also 
in the weak coupling limit. An interesting proposal in this direction 
has been made recently by Weinstein in 
Ref.~\cite{weinstein}, where, using the Contractor Renormalization Group 
method, he established the equivalence of 
various Hamiltonian free fermion theories with a class of generalized frustrated antiferromagnets.\\        
 
\noindent
{\bf Acknoledgments.}  We thank the organizers for the invitation to this 
workshop and for the stimulating 
and plesant atmosphere they helped 
to create. 
This research has been financed by grants from I.N.F.N. and M.U.R.S.T. . Through the years we greatly benefited from 
the many discussions with G. W. Semenoff.


\begin{thebibliography}{99}
\bibitem{neuberger} H. Neuberger, {\it Chiral fermions on the lattice}, talk delivered at Lattice99, june 29 - july 3, Pisa 
Italy, hep-lat/9909042 and references therein.
\bibitem{narayanan}R. Narayanan and H. Neuberger, Nucl. Phys. {\bf B443}, 305 (1995).
\bibitem{berruto1}F. Berruto, G. Grignani, G. W. Semenoff and P. Sodano, Phys. Rev. {\bf D59}, 034504 (1999); 
Annals of Phys. {\bf 275}, 254 (1999). 
\bibitem{berruto2}F. Berruto, G. Grignani and P. Sodano, hep-lat/9912038.
\bibitem{berruto3}F. Berruto, G. Grignani, G. W. Semenoff and P. Sodano, Phys. Rev. {\bf D57}, 5070 (1998). 
\bibitem{swieca}J. H. Lowenstein and J. A. Swieca, Ann. Phys. (N. Y.), {\bf 68}, 172 (1971); 
N. K. Nielsen and B. Schroer, Nucl. Phys. {\bf B120}, 62 (1977). 
\bibitem{zhitnitsky}A. R. Zhitnitsky, Phys. Lett. {\bf B165}, 405 (1985).
\bibitem{semenoff} G. W. Semenoff, Mod. Phys. Lett, {\bf A7}, 2811 (1992); E. Langmann and G. W. Semenoff, 
Phys. Lett. {\bf B297}, 175 (1992); M. C. Diamantini, E. Langman, G. W. Semenoff and P. Sodano, Nucl. Phys. 
{\bf B406}, 595 (1993).
\bibitem{gattringer}C. Gattringer and E. Seiler, Annals of Phys. {\bf 233}, 
97 (1994). 
\bibitem{grandou}T. Grandou, H.-T. Cho and H. M. Fried, Phys. Rev. {\bf D37},
 946 (1988).
\bibitem{coleman}S. Coleman, Commun. Math. Phys. {\bf 31}, 259 (1973).
\bibitem{thooft}G. `t Hooft, Nucl. Phys. {\bf B72},  461 (1974);
G. `t Hooft, Nucl. Phys. {\bf B75},  461 (1974).   
\bibitem{weinstein1}S. D. Drell, M. Weinstein and S. Yankielowicz, Phys. Rev. {\bf D14}, 11627 (1976).
\bibitem{smit} J. Smit, Nucl Phys. {\bf B175} 307 (1980).
\bibitem{weinstein}M. Weinstein, hep-lat/9910005; C. J. Morningstar and 
M. Weinstein, Phys. Rev. {\bf D54}, 4131, (1996); Phys. Rev. Lett. 
{\bf 73}, 1873 (1994).
\end{thebibliography}
\end{document}